\documentstyle[aps,multicol]{revtex}
\newcommand{\beq}{\begin{equation}} \newcommand{\eeq}{\end{equation}}
\newcommand{\bea}{\begin{eqnarray}} \newcommand{\eea}{\end{eqnarray}}
\newcommand{\bear}{\begin{eqnarray*}} \newcommand{\eear}{\end{eqnarray*}}
 
\newcommand{\rf}[1]{(\ref{#1})}

\begin{document}


\title {The  Bethe ansatz as a matrix product ansatz}

\author{Francisco C. Alcaraz  and Matheus J. Lazo }

\address {Universidade de S\~ao Paulo, Instituto de F\'{\i}sica de S\~ao 
Carlos, Caixa Postal 369, 13560-590 S\~ao Carlos, S\~ao Paulo, Brazil}

\date{\today}

\maketitle


\begin{abstract}

 The Bethe {\it ansatz} in its several formulations is the common tool for the 
 exact solution of one dimensional quantum Hamiltonians. This {\it ansatz}
 asserts that the several eigenfunctions of the Hamiltonians are given in 
 terms of a sum of permutations of plane waves. We present results that 
 induce us to expect that, alternatively, the eigenfunctions of all the 
 exact integrable quantum chains can also be expressed by a matrix 
 product {\it ansatz}. In this {\it ansatz} the several components of the
 eigenfunctions are obtained through the algebraic properties of properly 
 defined matrices. This {\it ansatz} allows an unified formulation of 
 several exact integrable Hamiltonians. We show how to formulate this 
 {\it ansatz} for a huge family of  quantum chains like the anisotropic 
 Heisenberg model,  Fateev-Zamolodchikov model,  Izergin-Korepin model, 
  Sutherland model,  $t-J$ model, Hubbard model, etc. 

\end{abstract}

\begin{multicols}{2} 

Since the pioneering work of Bethe in 1931\cite{bethe} 
the Bethe {\it ansatz} and its
generalizations proved to be a quite efficient tool in the description 
of the eigenvectors of a huge variety of one-dimensional quantum chains and two
dimensional transfer matrices (see e. g. \cite{baxter}-\cite{revschlo}  
for reviews). 
On the other hand in
the last  two decades \cite{affleck}-\cite{klumper} it has been shown that some special 
quantum chains, although not integrable through the Bethe {\it ansatz} have the 
components of its ground-state wavefunctions given in terms of a product 
of matrices. In this matrix product {\it ansatz} (MPA), apart from a normalization 
constant, these components are fixed by the algebraic properties of 
the matrices defining the MPA. 
In another context a 
MPA has also been applied quite successfully to the evaluation of the 
 stationary distribution of probabilities of some stochastic models 
 in one dimension \cite{derr1}. The time fluctuations of these stochastic models 
 are described by the ground state wavefunction 
 of a related spin Hamiltonian. The 
 simplest example is the one-dimensional asymmetric exclusion process
 \cite{derr1}
 whose related spin Hamiltonian is the anisotropic Heisenberg chain, or 
 XXZ chain, with appropriate boundary fields \cite{alcrit1}. The stationary 
 properties of the model are given in terms of the algebraic relations of the
 matrices appearing on the MPA. 
 This {\it ansatz} was used in a variety of problems including interface 
 growth \cite{krug}, boundary induced phase transitions \cite{derr1}, 
 \cite{derdomany}-\cite{derrevans}, the dynamics of shocks 
 \cite{derlebov} or traffic flow \cite{nagel}.

 An important development of the MPA that appeared in the context of 
 stochastic models is the dynamical matrix product {\it ansatz} (DMPA) 
 \cite{stinchshutz,reviewshutz}. This 
 {\it ansatz }  
 allows, whenever it is valid, the calculation of the probability 
 densities, of the stochastic system, at arbitrary times. In the related 
 spin Hamiltonian this DMPA asserts that not 
 only the ground-state wave function,
 as in the standard MPA, but an arbitrary wavefunction have components  
 expressed in terms of a matrix product {\it ansatz} whose matrices, in distinction
 of the standard MPA, are now time dependent.
 This DMPA was shown originally \cite{stinchshutz},\cite{sasamowada1} 
 to be valid for the problem of asymmetric diffusion 
of particles on the lattice. More  recently \cite{popkov} 
it was also shown that this 
 DMPA can also  be formulated in the context of stochastic models with 
 two species of hard-core particles. The condition of validity of the DMPA 
 reproduces the  subspace of  parameters where   the model is known 
 to be exactly 
 integrable. This fact induce us to conjecture that all the Hamiltonians, 
 related or not to stochastic models, that are solvable through the Bethe 
 {\it ansatz} may also be solvable by an  appropriate MPA. This would mean that the
 components of the eigenfunctions of the exact integrable models, that
 according 
 to the Bethe {\it ansatz}, are  normally 
  given by a combination of plane waves, can also be obtained from the 
 algebraic properties of the matrices defining the  MPA.

In this Letter we are going to show how the eigenspectra of  the exact 
integrable quantum chains, with global conservation laws, can be obtained by 
an appropriate matrix-product {\it ansatz}. In this way we were able to obtain 
the integrability of  several well known exact integrable models. Among the
models with one global conservation law  we have the XXZ chain \cite{yang}, 
the spin-$S$ 
Fateev-Zamolodchikov model \cite{fateev}, the Izergin-Korepin model 
\cite{izergin}, the solvable  spin-1 
model of Ref. \cite{alcbar1}, etc, and among the models with two global conservations we 
have the supersymmetric $t-J$ model \cite{schlo}, 
the spin-1 Sutherland  and Perk-Schultz models \cite{sutherland,perkshultz}, the
Hubbard model \cite{lieb}, as well the two-parameter integrable model  
presented in \cite{alcbar2}.
Moreover the matrix-product {\it ansatz} we propose enable us to show, with 
little effort, how to 
extend 
 the above mentioned models, by including 
 arbitrary range of hard-core interactions, without losing their exact 
 integrability. 

For brevity and in order to illustrate the proposed MPA we are going to 
present here  two examples. The  solution 
of the XXZ chain with arbitrary hard-core interactions among the up 
spins, as an example of a model with one global conservation law, 
and the solution
of the  
Hubbard model as an example of a model with two global conservations laws. 

The Hamiltonian of the XXZ chain with a hard-core  
exclusion of $S$ sites ($S=1, 2,
\ldots $) is the anisotropic Heisenberg chain (anisotropy $\Delta$) where any
two up spins are not allowed to occupy lattice sites at distances smaller than 
$S$. This Hamiltonian in a $L$-sites periodic chain is given by 
\beq \label{e1}
H_S = -{\cal P}_S \sum_{i=1}^L(\sigma_i^x\sigma_{i+1}^x + 
\sigma_i^y\sigma_{i+1}^y + \Delta\sigma_i^z\sigma_{i+1}^z){\cal P_S},
\eeq
where $\sigma^x, \sigma^y,\sigma^z$ are spin-$\frac{1}{2}$ Pauli 
matrices and ${\cal P}_S$ is a projector that projects out the configurations 
where any two up spins are at distances smaller than $S$. The case $S=1$ 
corresponds to the standard exactly solvable XXZ chain \cite{yang}. The 
conserved charge associated to the global 
conservation law  of (\ref{e1}) is the $z$- component of the total magnetization, 
or equivalently the  number of up spins (particles). The translation
invariance of the lattice also ensures that the momentum is also a good quantum
number. 

The {\it ansatz} we propose states that  any of the 
wavefunctions $|\psi_{n,P}>$ in the sector with $n$
spins up ($n=0,1,2,\ldots$) and  momentum $P = \frac{2\pi l }{L}$ 
($l=0,1,2,\ldots,L-1$) is given by a matrix product {\it ansatz}, i. e., their 
amplitudes are given by the trace of the  matrix product: 
\bea \label{e2}
&&|\psi_{n,P}> = \sum_{x_1,\ldots,x_n}^* \mbox{Tr} (E^{x_1-1}AE^{x_2-x_1-1}A \cdots
\nonumber \\ 
&&\cdots E^{x_n-x_{n-1}-1} A E^{L-x_n } \Omega_P) |x_1,\ldots,x_n>,
\eea
where $|x_1,\ldots,x_n>$ denotes the configurations with up spins 
at ($x_1,\ldots,x_n$) and the symbol ($*$) in the sum means the 
restriction to the configurations where $|x_{i+1}-x_i| \geq S$. The algebraic 
properties of the matrices $A$, $E$ and $\Omega_P$ will be fixed by the 
eigenvalue equation 
\beq \label{ea}
H_S|\psi_{n,P}> = \varepsilon_n |\psi_{n,P}>. 
\eeq
 The algebraic relations of the matrices $A$ and $E$ with $\Omega_P$ will fix
 the momentum of the eigenfunction.
The fact that $|\psi_{n,P}>$ has a momentum $P$ imply that the 
ratio of the amplitudes corresponding to  configurations $|x_1,\ldots,x_n>$ and 
($|x_1+1,\ldots,x_n+1>$ is $e^{-iP}$ and consequently from \rf{e2} we have 
the following  relations 
\beq \label{e22}
A\Omega_P = 
e^{-iP}\Omega_PA, \;   E\Omega_P = e^{-iP}\Omega_PA. 
\eeq
The eigenvalue equation \rf{ea} when applied to  the 
components of $|\psi_{n,P}>$ where all the up spins are at distances larger than $S$ give us the
constraint
\bea \label{e3}
&&\varepsilon_n 
{\mbox Tr} ( \cdots E^{x_i-x_{i-1}-1}A E^{x_{i+1}-x_i-1}A\cdots AE^{L-x_n}
\Omega_P)\nonumber \\
&& =-\sum_{i=1}^n [\mbox{Tr}( \cdots E^{x_i -x_{i-1}-2}A E^{x_{i+1}-x_i}A
\cdots AE^{L-x_n}\Omega_P) \nonumber \\
&& +\mbox{Tr} ( \cdots E^{x_1-x_{i-1}}AE^{x_{i+1}-x_i-2}A\cdots 
AE^{L-x_n}\Omega_P)] .
\eea
A convenient solution of this last equation is obtained by identifying the matrices $A$ 
as composed by $n$ spectral-parameter dependent matrices 
\beq \label{e4}
A = \sum_{j=1}^n  A_{k_j}E^{2-S},
\eeq
where the matrices $A_{k_j}$ obey the commutation relations 
\beq \label{e55}
EA_{k_j} = e^{ik_j}A_{k_j}E, \; A_{k_j}\Omega_P = e^{-iP(S-1)}\Omega_PA_{k_j},
\eeq
and   $k_j $ ($j =1,\ldots,n$) are in general complex numbers unknown {\it a 
priori} \cite{CO}. The energy and momentum  are given respectively by 
\beq \label{e5} 
\varepsilon_n = -\sum_{j=1}^n(e^{ik_j} + e^{-ik_j}), \; P = \sum_{i=1}^nk_j.
\eeq

The eigenvalue equation when applied to the other components of $|\psi_{n,P}>$ 
will give 
the commutation relations of the matrices $\{A_{k_j}\}$ among themselves. In 
fact this algebra is obtained from the components where any pair  of  up 
spins are
located at the closest positions ("matching conditions") 
$x_j$ and $x_{j+1}= x_j +S$, namely
\bea \label{e6}
&&A_{k_j}A_{k_l} = S(k_j,k_l)A_{k_j}A_{k_l} \; (i\neq j),\;  A_{k_j}^2=0 ,
\nonumber \\
&&S(k_j,k_l) = - \frac{e^{i(k_j+k_l)} +1 -2\Delta e^{ik_j}} 
{e^{i(k_j+k_l)}+1-2\Delta e^{ik_l}} , \; i,j=1,\ldots,n.
\eea
No new algebraic relations appear for the matrices $A_{k_j}$ and the 
associativity of the algebra \rf{e55} and \rf{e6} follows from property 
$S(k_i,k_j)S(k_j,k_i)=1$.
The cyclic property of the trace in (\ref{e2}), together with the algebraic
relations (\ref{e6}) give us, for each $k_j$ (j =1,\ldots,n),   the conditions 
\bea \label{e7} 
&&\mbox{Tr}(A_{k_1}\cdots A_{k_n}E^{L-n(S-1)}\Omega_P) = 
e^{-ik_j(L-nS +n)} e^{-iP(S-1)} \nonumber \\
&&
\times (\prod_{l=1}^n 
S(k_j,k_l)) \mbox{Tr}(A_{k_1}\cdots A_{k_n}E^{L-n(S-1)}\Omega_P),
\nonumber
\eea
that fix the up to now free complex spectral parameters $\{k_j\}$ 
\beq \label{e8}
e^{ik_jL}=\prod_{l=1}^n S(k_j,k_l)e^{i(k_j-k_l)(S-1)}, \ (j=1,\ldots,n).
\eeq
The wavefunctions,  obtained by using (\ref{e2}) \rf{e5} and (\ref{e6})  can 
be written as a combination of plane waves ("wave numbers" $\{k_j\}$), 
and apart from a normalization
constant coincides with those obtained by the standard  Bethe {\it ansatz} 
\cite{yang}
 for $S=1$ or arbitrary values of $S$ \cite{alcbar3}.


As a second example we consider the standard Hubbard model whose 
Hamiltonian in a periodic lattice with $L$ sites is given by  
\beq \label {e9}
H_h = -t\sum_{j,\sigma } (c_{j,\sigma}^+c_{j+1,\sigma} + 
c_{j+1,\sigma}^+c_{j,\sigma})+u\sum_{j} \eta_{j,-} \eta_{j,+},
\eeq
where $c_{j,\sigma}^+$ are creation operators of electrons of 
spin $\sigma=\pm $ at site $j$, and $\eta_{j,\sigma} = \sum_{ja} 
c_{j,\sigma}^+c_{j,\sigma}$
 are the number  operators of electrons of spin $\sigma$ at the  site $j$. 
 In this case we have two global conservation laws $n_{\pm}$, 
 corresponding to the 
 number of electrons with spin $\sigma = \pm$, and the toal number of 
 electrons is $n = n_+ + n_-$. In order to form our 
 MPA we associate, as before,  the  matrix $E$ to the empty 
 sites, the matrices $X^+$, $X^-$ to the single occupied sites with electrons 
 with spin up and down, respectively,  and the matrix 
 $X^0 \equiv X^+E^{-1}X^-$ to the sites with 
 double occupancy. Our MPA asserts that the components of the 
 eigenfunction $|\psi_{n,P}>$ of energy 
 $\epsilon_n$ and momentum $P = 2\pi l /L$ ($l=0,1,\ldots,L-1$) 
 corresponding to the configuration 
 $|x_1,Q_1; \ldots, x_n,Q_n>$ where the non-empty sites ($x_1,\ldots,x_n$) 
 have occupation ($Q_1,\ldots,Q_n$) ($Q_i=+,-,0$) will be given by the trace
 \beq \label {e10}
 \mbox{Tr} (E^{x_1-1}X^{Q_1}E^{x_2-x_1-1}X^{Q_2}\cdots X^{Q_n}E^{L-x_n}\Omega_P).
 \eeq
 The momentum of the state $P$, as in \rf{e22},  is 
 fixed by imposing the commutation 
 relations $X^{Q}\Omega_P = e^{-iP}\Omega_PX^Q$, ($Q =+,-,0$) and
 $E\Omega_P = e^{-iP}\Omega_PE$. The eigenvalue equation $H_h |\Psi_{n,P}> =
 \varepsilon_n |\Psi_{n,P}>$ will provide the algebraic relations of the 
 matrices $E$ and $X^Q$. 
 
 The components of the wavefunction corresponding 
 to the configurations where all the particles are at distances 
 $|x_{i+1} - x_i| >1$ will give a generalization of (\ref{e2}) whose solution 
 is obtained by introducing the convenient $n$ 
 spectral-parameter dependent matrices
 \beq \label{e11}
 X^Q = \sum_{j=1}^nEX_{k_j}^Q, \; (Q = +,-) ,
 \eeq
 whose commutation relations with the matrices $E$ and $\Omega_P$ are 
 \beq \label{e12}
  EX_{k_j}^Q = e^{ik_j}X_{k_j}^QE, \; X_{k_j}^Q \Omega_P= \Omega_PX_{k_j}^Q . 
 \eeq 
 The energy and momentum  in terms of these unknown complex 
 spectral parameters are
 given, as in \rf{e5},  by 
 \beq \label{e13} 
 \varepsilon_n = -\sum_{j=1}^n(e^{ik_j} + e^{-ik_j}), \; \; P =\sum_{j=1}^n 
 k_j.
 \eeq

 The components where the particles occupy the closest  positions and 
 those where 
 we have double occupancy, give us, by using (\ref{e12}) and (\ref{e13}) the 
 algebraic relations 
 \bea \label {e14}
 && X_{k_l}^QX_{k_j}^Q = S_{QQ}^{QQ}(k_l,k_j)X_{k_j}^QX_{k_l}^Q, \; \; 
 (X_{k_j}^Q)^2 = 0, \nonumber \\ && X_{k_l}^QX_{k_j}^{Q'} =
 S_{QQ'}^{QQ'}(k_l,k_j)X_{k_j}^{Q'}X_{k_l}^Q +
 S_{Q'Q}^{QQ'}(k_l,k_j)X_{k_j}^{Q}X_{k_l}^{Q'} \nonumber  \\ &&S_{QQ}^{QQ}= -1,
 \;  S^{QQ'}_{Q'Q}(k_l,k_j) = -ue^{i(k_l +k_j)}/\alpha, \nonumber \\  &&
 S_{QQ'}^{Q'Q}(k_l,k_j)=t(e^{ik_l} -e^{ik_j}) (1+e^{i(k_l+k_j)})/\alpha,
 \nonumber \\ &&\alpha = ue^{i(k_l+k_j)} + t(1+e^{i(k_l+k_j)})(e^{ik_l}
 -e^{ik_j}), 
 \eea 
 where $Q' = -Q$ and  $Q= \pm$.  The relations (\ref{e12}) and
 \rf{e14} define completely the
algebra whose structural constants are the well know $S$-matrix of the Hubbard
model \cite{lieb}. Since the several components of the wavefunction should be uniquely
related the above algebra should be associative. This associativity imply
that the above $S$-matrix should satisfy the Yang-Baxter relations 
\cite{yang2,baxter}, which 
is indeed the case \cite{lieb}. The components of the wavefunction corresponding to 
the configurations where we have three or  four  particles 
in next-neigboring sites  
would give in principle new relations involving three  or four 
operators $X_{k_j}^Q$. 
These new relations are however consequences of the above relations 
\rf{e12} and \rf{e14}. 
The cyclic property of the trace in \rf{e10} and the algebraic relations 
 \rf{e12} and \rf{e14} will imply 
 \bea \label{en1}
 &&\mbox{Tr} (X_{k_1}^{Q_1} \cdots X_{k_{j-1}}^{Q_{j-1}}X_{k_j}^{Q_j} \cdots 
 X_{k_n}^{Q_n} E^{L} \Omega_P) = (-1)^n e^{ik_jL}   \nonumber \\ 
 && \sum_{Q'_1,\ldots,Q'_n} {\cal T}(\{Q\},\{Q'\}) 
 \mbox {Tr} (X_{k_1}^{Q'_1} \cdots X_{k_n}^{Q'_n} E^{L} \Omega_P). 
 \eea
 where
 \beq 
 {\cal T}(\{Q\},\{Q'\}) = \sum_{Q"_1,\ldots,Q"_n} 
 \prod_{i=1}^n S_{Q'_i,Q"_i}^{Q_i,Q"_{i+1}} \nonumber
 \eeq
is the  transfer matrix of a non-homogeneous six-vertex model 
defined on a cylinder of perimeter $n$ and with Boltzman weights given 
by the $S$-matrices defined in \rf{e14}. 
 The eigenvalues of this auxiliary problem $\Lambda_n(k_i;k_1,\ldots,k_n)$ can be obtained in a 
 standard way by 
 the  coordinate-Bethe {\it ansatz} \cite{yang}
or by the quantum inverse scattering method \cite{kulish}. Using these 
eigenvalues in  the relation \rf{en1} the 
spectral parameters $\{k_j\}$ will be fixed by the solutions of the system of
equations 
\beq \label{e15}
e^{-ik_jL} = (-1)^n \Lambda_n(k_j;k_1,\ldots,k_n), \; (j=1,\ldots,n).
\eeq
These last equations coincide with the Bethe-{\it Ansatz} equations derived 
through the standard coordinate Bethe {\it ansatz} \cite{lieb}.

Generalizations of  our MPA is quite simple to implement \cite{TB}. For example the
 solution of the 
excluded volume 
 Hubbard model where  electrons with spin up (down) exclude other electrons 
at $S_+$ ($S_-$) sites on its right, but allows double occupancy at any site, 
is just 
obtained by changing in \rf{e11} $X^Q = \sum_j E^{S_Q +1} X_{k_j}^Q$ 
($Q=\pm$). 

We 
have also shown \cite{TB} that the above MPA also works for the other known exact 
integrable models with two conserved global quantities ($U(1)\otimes U(1)$), 
like  
the Essler-Korepin-Schoutens model \cite{essler2}
or the generalized two-parameter integrable model introduced in \cite{alcbar2}. 
 In those last cases the same MPA presented above for the Hubbard 
model apply except that now the matrices $X^0$ associated to the sites with 
double occupancy  
are given in terms of new  operators $X^0 = Y^+E^{-1}Y^-$, and we 
were able to rederive all the results obtained for  those models through the standard 
Bethe {\it ansatz}. Hamiltonians  with two global conservations 
that do  not allow 
double 
occupancy like the stochastic Hamiltonian associated to the asymmetric 
diffusion with two kinds of particles \cite{alcbar3}, or the supersymmetric  $t-J$ quantum
chain \cite{schlo}
or the  $SU(3)$ Sutherland and Perk-Schultz models 
\cite{sutherland,perkshultz}, 
are obtained through the MPA \rf{e10}, without the matrices $X^0$.  

Models of spin one with a single global conservation law (conservation of the 
$z$-magnetization) like the Fateev-Zamolodchikov model \cite{fateev}, the
Izergin-Korepin model \cite{izergin}, or the solvable spin-1 Hamiltonian introduced 
in \cite{alcbar2}, have their
solutions given by a MPA similar as in \rf{e2}, where now we associate  the 
matrices $E$, $A$ and $BE^{-1}B$ to the sites occupied by particles 
with spin $S^z = -1$, 0 or +1, respectively.

   We have also obtained an appropriate 
    extension of the presented  MPA  to non periodic, but 
   exact integrable boundaries as,  for example, the 
   XXZ chain with surface fields.

It is interesting to notice  that in the cases of exact integrable 
Hamiltonians associated to stochastic models, 
as in \cite{alcrit1,reviewshutz}, since we can write 
 all eigenfunctions in a MPA, our results imply that  we can equivalently write at any time  
 the probability distribution  of the model in terms of a time-dependent 
 MPA, as happens in the DMPA \cite{stinchshutz}. 

 In conclusion, we have shown that the  eigenfunctions of a huge variety
 of exact integrable quantum chains can be expressed  on an unified way 
 in terms of a matrix product ansatz, whose matrices satisfy an associative
 algebra.  The associativity of the algebra that warranties the exact 
 integrability of the model is a 
 consequence of the Yang-Baxter relations.

This work has been partially supported by  FAPESP, CNPq and CAPES  (Brazilian agencies).

\end {multicols}

\begin{thebibliography}{99}

\bibitem{bethe}H. A. Bethe, Z. Phys. {\bf 71}, 205 (1931). 

\bibitem{baxter} R. J. Baxter, {\it Exactly solved models in statistical
mechanics}
    (Academic Press, New York, 1982).
	
\bibitem{revkore} 
  V. E. Korepin, A. G .Izergin and N. M. Bogoliubov, {\it Quantum Inverse
Scattering Method, Correlation Functions and Algebraic Bethe Ansatz}
(Cambridge University Press, Cambridge, 1992). 

\bibitem{revessler}
 F. H. L. Essler and  V. E. Korepin, {\it Exactly Solvable Models of Strongly
 Correlated Electrons} (World Scientific, Singapore, 1994).

 \bibitem{revschlo}
 P. Schlottmann, Int. J. Mod. Physics B, {\bf 11}, 355 (1997).
 
 \bibitem{affleck} 
 I. Affleck, T. Kennedy, E. H. Lieb and H. Tasaki, Commun.
Math. Phys.
 {\bf 115}, 477 (1988).

 \bibitem{arovas} D. P. Arovas, A. Auerbach, F. D. M. Haldane, Phys. Rev. Lett.
 {\bf 60},  531 (1998). 

 \bibitem{fannes}
 
M. Fannes, B. Nachtergaele, R. F. Werner, Commun. Math. Phys. {\bf 144}, 443 
(1992).

\bibitem{klumper} 
 A. Kluemper, A. Schadschneider and J. Zittarz, Z. Phys. B {\bf 87}, 281
 (1992); 
Europhys. Lett. {\bf24}, 293 (1993).

\bibitem{derr1}
B. Derrida, M. R. Evans, V. Hakim and V. Pasquier, J. Phys. A {\bf 26}, 1493 
(1993). 

\bibitem{alcrit1} F. C. Alcaraz, M. Droz, M. Henkel,
and V. Rittenberg,
Ann. Phys. (N.Y.) {\bf 230}, 250 (1994).

\bibitem{krug}
J. Krug and H. Spohn, {\it Solids Far From Equilibrium} Eds. C. Godr\`eche 
(cambridge University Press, Cambridge, 1991).

\bibitem{derdomany}
B. Derrida, E. Domany and D. Mukamel, J. Stat. Phys. {\bf69}, 667 (1992).

\bibitem{domashutz}
R. Domany and E. Sch\"utz, J. Stat. Phys. {\bf 72}, 277 (1993).

\bibitem{derrevans} 
M. R. Evans, D. P. Foster, C. Godr\`eche and D. Mukamel, J. Stat. Phys. 
{\bf 80}, 69 (1995).

\bibitem{derlebov}
B. Derrida, J. L. Lebowitz and E. R. Speer, J. Stat. Phys. {\bf 89}, 135 (1997).

\bibitem{nagel} 
K. Nagel and M. Schreckenberg, J. Physique {\bf 2}, 2221 (1992)

\bibitem{stinchshutz}
R. B. Stinchcombe and G. M. Sch\"utz, Phys. Rev. Lett. {\bf 75}, 140 (1995);
Europhys. Lettt. {\bf 29}, 663 (1995). 

\bibitem{sasamowada1}
T. Sasamoto and M. Wadati, J. Phys. Soc. Japan {\bf 66}, 2618 (1999).

\bibitem{popkov} V. Popkov, M. E. Fouladvand and G. M. Sch\"utz, 
J. Phys. A {\bf 35}, 7187 (2002).

\bibitem{yang} C. N. Yang and C. P. Yang, Phys. Rev. {\bf 150}, 321 (1966).

\bibitem{fateev} 
A. B. Zamolodchikov and V. Fateev, Sov. J. Nucl. Phys. {\bf 32}, 298 (1980).

\bibitem{izergin}
A. G. Izergin and V. E. Korepin, Commun. Math. Phys. {\bf 79}, 303 (1981).

\bibitem{alcbar1}
F. C. Alcaraz and R. Z. Bariev, J. Phys. A {\bf 34}, L467 (2001).

\bibitem{schlo}
 P. Schlottmann, Phys. Rev. B {\bf 36}, 5177 (1987).

 \bibitem{sutherland}
 B. Sutherland, Phys. Rev. B {\bf 12}, 3795 (1975).

 \bibitem{perkshultz}
 J. H. H. Perk and C. L. Schultz, Phys. Lett. A {\bf 84}, 407 (1981).

 \bibitem{lieb}
 E. H. Lieb and F. Y. Wu, Phys. Rev. Lett. {\bf 20}, 1445 (1968).

 \bibitem{alcbar2}
 F. C. Alcaraz and R. Z. Bariev, J. Phys. A {\bf 32}, L483 (1999).
\bibitem{reviewshutz} G. M. Sch\"utz, in "Phase Transitions and Critical
Phenomena", vol. 19, Edited by  C. Domb and J. Lebowitz (Academic, London,
2000)

\bibitem{CO}
Actually the most general relation $A = \sum_j E^{\alpha}A_{k_j}E^{\beta}$ 
 with ($\alpha, \beta \in Z$) could be used. However
  \rf{e4} is more convenient since the S-matrix in \rf{e6} will be 
  independent from the hard-core size of the up spins.

\bibitem{alcbar3}
F. C. Alcaraz and R. Z. Bariev, in {Statitical Physics in the Eve of the 
 21st Century}, edited by M. T. Batchelor and L. T. Wille, Series on 
 advances in Statistical Mechanics, Vol. 14 (World Scientic, Singapore, 1999),
 ({\it Preprint} cond-mat/9904042).

 \bibitem{yang2}
 C. N. Yang, Phys. Rev. Lett. {\bf 19}, 1312 (1968).

 \bibitem{kulish}
 P.P. Kulish and N. Yu. Reshetikhin, Sov. Phys. -JETP {\bf 53}, 108 (1981). 

 \bibitem{TB} 
 F. C. Alcaraz and M. J. Lazo, to be published.

 \bibitem{essler2} F. H. L. Essler, V. E. Korepin and K. Schoutens, Phys. Rev.
 Lett. {\bf 68}, 2960 (1992); Phys. Rev. Lett. {\bf 70}, 73 (1993).

 \bibitem{alcbar4}
 F. C. Alcaraz and R. Z. Bariev, Braz. J. Phys. {\bf 30}, 13 (2000).

 

\end{thebibliography}
\end{document}